\documentclass[aps,twocolumn,PRB,reprint,superscriptaddress]{revtex4}

\usepackage{amsfonts}
\usepackage{amsmath}
\usepackage{amssymb}
\usepackage{graphicx}
\usepackage{color}
\usepackage{tensor}
\usepackage{upgreek}
\usepackage{accents}
\usepackage[percent]{overpic}
\usepackage{times}
\usepackage{subfigure}
\usepackage{latexsym}
\usepackage{bm}
\usepackage{anyfontsize}
\usepackage{hyperref}
\usepackage{multirow}
\usepackage{booktabs}
\usepackage{makecell}

\begin{document}

\title{Quantum Transport on the Surfaces of Topological Nodal-line Semimetals}
\author{Jun-Jie Fu}
\affiliation{National Laboratory of Solid State Microstructures, Department of Physics, Nanjing University, Nanjing 210093, China}
\author{Shu-Tong Guan}
\affiliation{National Laboratory of Solid State Microstructures, Department of Physics, Nanjing University, Nanjing 210093, China}
\author{Jiao Xie}
\affiliation{National Laboratory of Solid State Microstructures, Department of Physics, Nanjing University, Nanjing 210093, China}
\author{Jin An}
\email{anjin@nju.edu.cn}
\affiliation{National Laboratory of Solid State Microstructures, Department of Physics, Nanjing University, Nanjing 210093, China}
\affiliation{Collaborative Innovation Center of Advanced Microstructures, Nanjing University, Nanjing 210093, China}

\begin{abstract}
Topological nodal-line semimetals are always characterized by the drumhead surface states at the open boundaries. In this paper we first derive an analytical expression for the surface Green's function of a nodal-line semimetal. By making use of this result, we explore the charge and spin transport properties of a metallic chain on the surface of a nodal-line semimetal, as functions of the gate voltage applied on the top of the material. According to the size of the nodal loop, due to the coupling to the surface modes, the charge conductance in the chain is found to show a robust plateau at $e^{2}/h$, or to exhibit multiple valleys at $e^{2}/h$. Correspondingly, the spin polarization of the transmitted current is $100\%$ at the plateau region, or exhibits multiple peaks at nearly $100\%$. This feature can be viewed as a transport signature of the topological nodal-line semimetals.
\end{abstract}

\date{\today}

\maketitle

\section{INTRODUCTION}
As one of topological semimetal materials, topological nodal-line semimetals (NLSs) have been attracting great interest in condensed matter physics\cite{PhysRevB.84.235126,PhysRevB.90.115111,PhysRevB.90.205136,Xie10.1063,PhysRevLett.115.036806,PhysRevLett.115.036807,acs.nanolett.5b02978,PhysRevLett.115.026403,PhysRevB.92.081201,JPSJ.85.013708,PhysRevB.93.205132,PhysRevLett.117.096401,PhysRevLett.117.087402,RevModPhys.88.035005,PhysRevB.94.195104,PhysRevLett.116.127202,PhysRevB.93.121113,ncomms10556,ncomms11696,PhysRevB.93.201104,PhysRevLett.117.016602,PhysRevB.94.121108,Hirayama2017,Yu2017,PhysRevLett.118.176402,PhysRevB.95.235104,PhysRevB.95.205134,PhysRevB.96.081107,PhysRevB.96.155206,PhysRevB.95.125126,PhysRevB.96.161105,Yang23746149,PhysRevB.98.245422,PhysRevLett.120.146602,PhysRevX.8.031044,PhysRevB.98.041105,Takane2018,LI2018535,PhysRevB.98.121111,sciadvaau6459,PhysRevB.100.125136,PhysRevLett.122.186801,PhysRevB.100.115101,acsaelm.9b00061,science.aav2327,PhysRevLett.124.056402,PhysRevX.10.011026,PhysRevB.104.045143,PhysRevB.103.125131,RevModPhys.93.025002,PhysRevX.11.031017,PhysRevB.103.L161109,feng2022,PhysRevB.105.235103,PhysRevB.105.125120,PhysRevB.105.214434,PhysRevB.107.205411,PhysRevB.107.085139}. Topological NLSs have one-dimensional (1D) band crossings between conduction and valence bands in the momentum space, and always hold characteristic flat band at certain boundary: drumhead surface states, exhibiting topologically distinct new features. According to the degeneracies of band crossings, they can be classified into the Weyl or Dirac NLSs, which are doubly or fourfold degenerate respectively. The topological structures of the 1D band crossings support various possibilities such as
nodal-net\cite{PhysRevLett.115.036807,PhysRevB.92.045108,PhysRevX.8.031044,PhysRevMaterials.2.014202}, nodal-chain\cite{nature19099,PhysRevLett.119.036401,PhysRevLett.119.156401,jpclett8b02204,C8CP02810A,yanqing2018,Wu_2023}, and Hopf-link \cite{PhysRevLett.121.106403,PhysRevB.96.041102,PhysRevB.96.041103,PhysRevB.96.041202,PhysRevB.96.081114,PhysRevB.96.201305,PhysRevB.97.155140,lian2019} NLSs, most of which have been realized in experiments.

\begin{figure}[ht]
  \begin{center}
	\includegraphics[width=8.5cm,height=5.95cm]{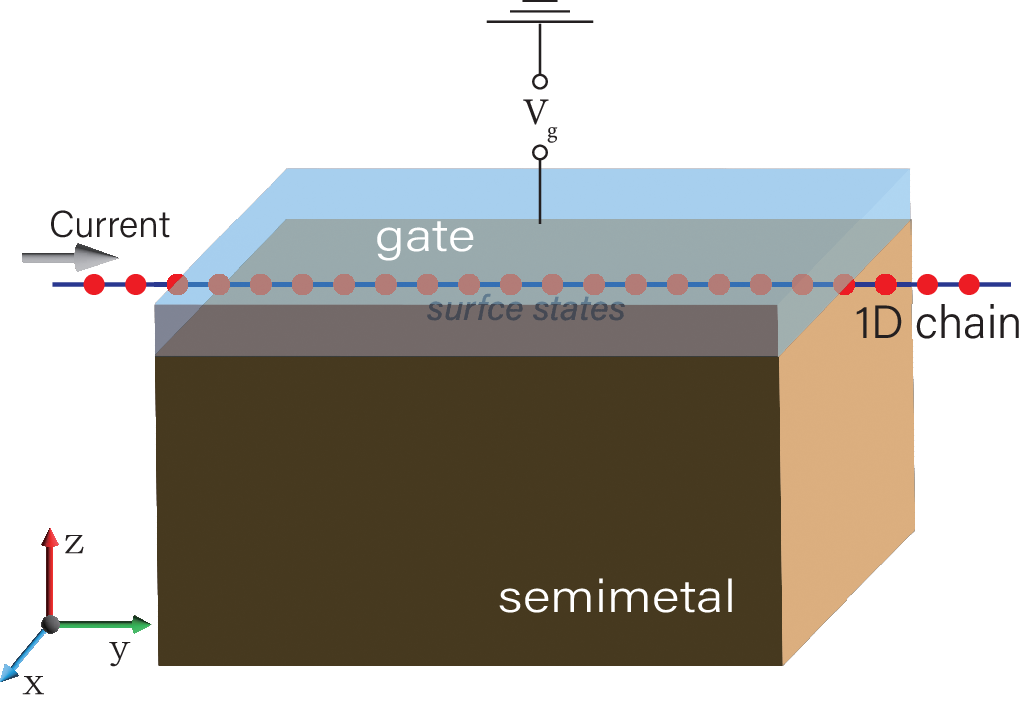}
  \end{center}
  \vspace{-0.2cm}
  \caption{ Device designed to explore the quantum transport of a 1D metallic chain on the surface of a topological semimetal, where a gate voltage is applied on the surface to modify the chemical potential.
  }
\label{fig1}
\end{figure}

To date, lots of materials have been theoretically proposed to be NLSs\cite{Xie10.1063,PhysRevLett.115.036806,PhysRevLett.115.036807,JPSJ.85.013708,PhysRevB.93.205132,PhysRevLett.117.096401,PhysRevB.94.195104,PhysRevB.93.121113,Hirayama2017,Yu2017,PhysRevB.95.235104,PhysRevB.96.155206,PhysRevB.98.245422,PhysRevLett.121.106403,PhysRevB.100.115101,PhysRevB.104.045143,PhysRevB.103.L161109,PhysRevB.107.205411}, and with the help of angle-resolved photoemission spectroscopy and quantum oscillations, a few NLS materials have also been experimentally verified\cite{ncomms10556,ncomms11696,PhysRevB.93.201104,PhysRevLett.117.016602,PhysRevB.94.121108,PhysRevB.95.205134,PhysRevB.95.125126,PhysRevX.8.031044,PhysRevB.98.041105,Takane2018,LI2018535,PhysRevB.98.121111,sciadvaau6459, PhysRevB.100.125136,acsaelm.9b00061,science.aav2327,PhysRevLett.124.056402,PhysRevX.10.011026,PhysRevB.103.125131,PhysRevB.105.125120,PhysRevB.105.214434}. These NLSs have been found or proposed to reveal much more exciting features. These include non-Abelian band topology\cite{quansheng2019,PhysRevB.101.195130,PhysRevLett.129.263604}, distinct collective modes \cite{PhysRevB.93.085138,PhysRevB.104.245301}, correlation effects\cite{PhysRevB.93.035138,shaonp}, second-order NLSs\cite{PhysRevLett.125.126403,PhysRevLett.128.026405,PhysRevB.107.035128}, and novel transport, the last of which is also exhibiting exotic phenomena such as flat Landau-level spectrum\cite{PhysRevB.92.045126}, anomalous Hall current\cite{PhysRevLett.118.016401,WOS:000548310100001,PhysRevLett.130.166702,PhysRevB.107.125138}, parity anomaly\cite{PhysRevB.97.161113,PhysRevB.98.155125,PhysRevResearch.2.043311}, fully spin-polarized transport\cite{PhysRevB.101.075125}, anomalous Andreev reflection\cite{PhysRevB.101.094508}, and weak antilocalization\cite{PhysRevLett.122.196603, kim2022}. However, less attention has been paid on the transport phenomenon directly related to the drumhead surface states of the NLSs, and one may ask a question: Whether do there exist transport properties characterized by these surface modes? To answer this question, in this paper, we study the quantum transport of a 1D metallic chain deposited on the top surface of a Weyl NLS with a single nodal loop, as shown in Fig. \ref{fig1}. The chain is expected to be coupled effectively to the surface modes. We found that the charge conductance would exhibit half transmission phenomenon: the charge conductance forms a robust plateau of $e^2/h$, while the spin polarization of the transmitted current would reach $100\%$.

This paper is organized as follows. In Sec. \ref{sect1}, we introduce our model of the topological NLSs and then we give our analytical result of the surface Green's function for the 1D effective model. In Sec. \ref{sect2}, based on the derived Green's function, we discuss successively the quantum transport properties of the metallic chain, which is coupled with the lower dimensional counterparts of our model, and then we turn back on the discussion of the chain on the surface of a topological NLS. In Sec. \ref{sect3}, we summarize the results.

\section{SURFACE GREEN'S FUNCTION OF THE 1D EFFECTIVE MODEL}
\label{sect1}

The minimal lattice model of the topological semimetals we study is given by:
\begin{equation}
	\begin{split}
		\mathcal{H}=&(m-\cos k_x-\cos k_y-\cos k_z)\sigma_x\\
		&+\eta \sin k_y\sigma_y+\lambda \sin k_z\sigma_z,
	\end{split}
 \label{eq1}
\end{equation}
where $\bm{\sigma}=(\sigma_{x},\sigma_{y},\sigma_{z})$ are spin Pauli matrices and $m$, $\eta$, $\lambda$ are adjustable parameters. When $m$ is properly chosen to be within $1<m<3$, this minimal model describes the topologically nontrivial semimetals. If $\eta=0$, this model depicts a NLS with a nodal loop located at $k_z=0$ and $\cos k_x+\cos k_y=m-1$, which is protected by mirror symmetry. This NLS is characterized by the zero-energy drumhead surface states when the system is open along $z$ direction. If $\eta\neq0$, this model describes a $\mathcal{T}$-breaking Weyl semimetal with a pair of Weyl points located at $\mathbf{k}=(\pm \cos^{-1}(m-2),0,0)$. When the boundary perpendicular to $z$ direction is open, this Weyl semimetal has a surface edge band: $E_{\text{arc}}(k_x,k_y)=-\text{sgn}(\lambda)\eta\sin k_y$. This edge band is restricted within the closed loop: $\cos k_x+\cos k_y=m-1$, at which the edge band is touching the bulk bands.

The deposited 1D metallic chain can still be viewed as a quasi-1D chain since the coupling to the surface modes can be taken into account by the self-energy. This self-energy relays mainly on the surface Green's function of the topological semimetal. So in the following, we first study the surface Green's function of the corresponding 1D effective model and then we apply the results to the transport in the metallic chain.

\begin{figure}[ht]
  \begin{center}
	\includegraphics[width=8.5cm,height=3.64cm]{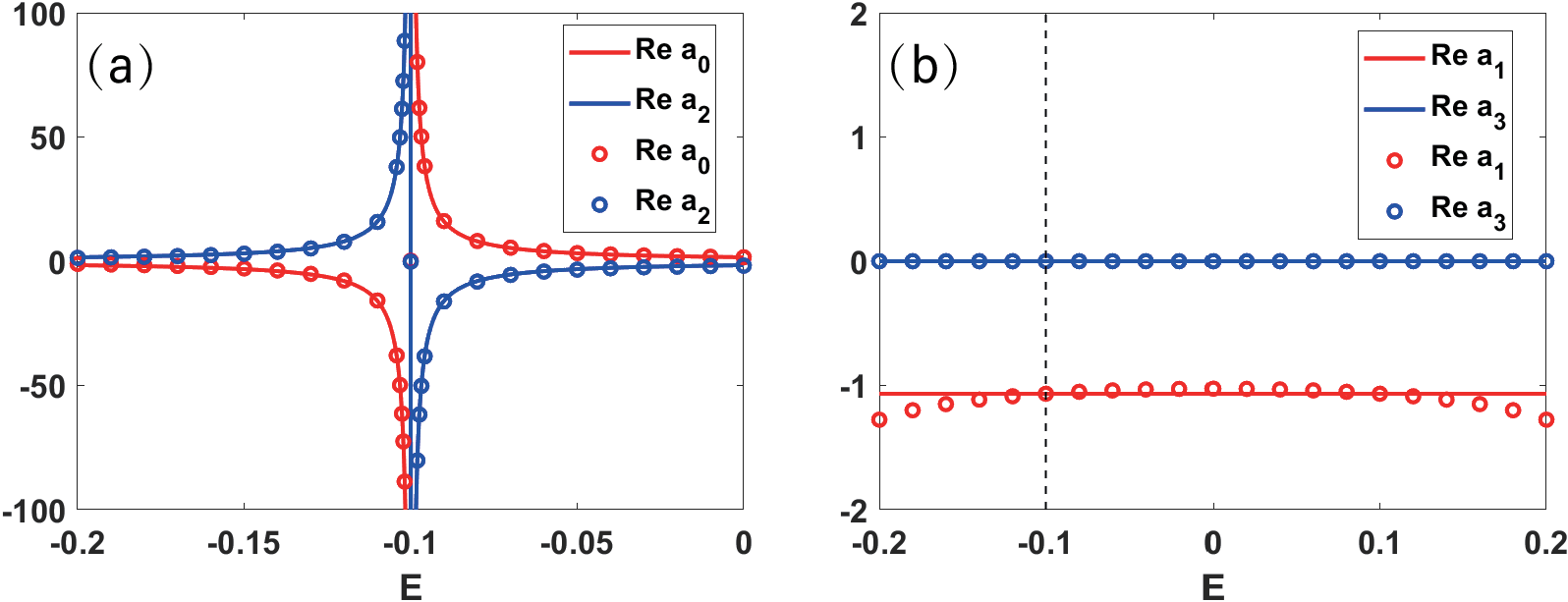}
  \end{center}
  \vspace{-0.2cm}
  \caption{ Real parts of the expansion coefficients of the surface Green function $g^r(E)$ near the singularity $E=-\Delta$, when $\xi=0.8$, $\lambda=0.5$ and $\Delta=0.1$, where the solid lines (open circles) represent the analytical (numerical) results.
  } \label{fig2}
\end{figure}

The 1D effective model is actually the 1D version of our minimal model with $k_{x}$ and $k_{y}$ being fixed as good quantum numbers:
\begin{equation}
	\mathcal{H}_{\text{eff}} (k_{z})=(\xi-\cos k_{z})\sigma_x+\Delta\sigma_y+ \lambda \sin k_{z}\sigma_z,
 \label{eq2}
\end{equation}
where $\xi=m-\cos k_{x}-\cos k_{y}$ and $\Delta=\eta\sin k_{y}$ can be viewed as independent parameters in this effective model. One can easily check that an open chain of this effective model possesses an end mode with energy $E=-$sgn$(\lambda)\Delta$ (sgn$(\lambda)\Delta$) at left (right) end only if $|\xi|<1$. Consider a semi-infinite system of this effective model, where its open boundary is located at $z=0$. The retarded surface Green's function $g^r$ obeys:
\begin{equation}
	(g^r)^{-1}+t g^r t^{\dag}=E^{+}-h_0,
	\label{relationship}
\end{equation}
where $h_0=\xi \sigma_x + \Delta \sigma_y$, $t=-\frac{1}{2}(\sigma_x-i\lambda \sigma_z)$, and $E^{+}=E+i0^{+}$. When $E\to-\text{sgn}(\lambda)\Delta$, the expression of $g^r(E)$ can be obtained analytically:
\begin{widetext}
\begin{equation}
	\begin{split}
		g^r(E) =
		\begin{cases}
			\frac{2 \vert\lambda\vert (1-\xi^2)}{(1+|\lambda|)^2}\{1-\text{sgn}(\lambda) \sigma_y\}\frac{1}{E^{+}+\text{sgn}(\lambda)\Delta}-\frac{2\xi}{1+|\lambda|}\sigma_x, & |\xi|<1 \\
			\\
			\frac{-2|\lambda|i}{(1+|\lambda|)^2}\{1-\text{sgn}(\lambda) \sigma_y\}-\frac{2}{1+\vert\lambda\vert}\sigma_x, & |\xi|=1,\Delta=0 \\
			\\
			\frac{\sqrt{8\Delta\lambda|\lambda|}}{(1+|\lambda|)^2\sqrt{E^{+}+\text{sgn}(\lambda)\Delta}}\{1-\text{sgn}(\lambda) \sigma_y\}-\frac{2}{1+\vert\lambda\vert}\sigma_x, & |\xi|=1,\Delta \neq 0 \\
			\\
			\frac{\Delta}{S+(1+|\lambda|)^{2}/4}\{-\text{sgn}(\lambda)+\sigma_{y}\}-\frac{1}{\sqrt{-S}}\sigma_{x}
+\frac{1}{(1-\xi^2)S}\{(S+\frac{1+\lambda^2}{4})+\frac{\lambda}{2}\sigma_y\}\{E^{+}+\text{sgn}(\lambda)\Delta\}, & |\xi|>1
		\end{cases}
	\end{split}
\label{eq4}
\end{equation}
\end{widetext}
where $S=-\frac{1}{4}(\xi+\sqrt{\xi^2+\lambda^2-1})^2$ and the value of the square root in the numerator of the first term in the third line is $-i\sqrt{8\vert\Delta\vert}\vert\lambda\vert$ if $\lambda\Delta<0$. For the details of its derivation, see Appendix \ref{a1}. This surface Green's function explicitly exhibits the existence of the end mode at $E=-$sgn$(\lambda)\Delta$ if $|\xi|<1$. In Fig. \ref{fig2}, we show for $|\xi|<1$ both the analytical and the corresponding numerical results of the expansion coefficients of $g^r(E)=a_{0}(E)+(a_{1}(E),a_{2}(E),a_{3}(E))\cdot\bm{\sigma}$ near the singularity. They agree quite well with each other.

With $g^r$, we are ready to explore the quantum transport of the metallic chain deposited on the surface of a 3D NLS. But we notice that the quantum transport of the chain which is coupled to the corresponding 1D or 2D effective system is also very meaningful and is found to exhibit characteristic behaviors closely related to the 3D case. So in the following section we will begin with these two lower dimensional cases and then we return back to the transport on the surface of a 3D NLS system.

\section{Charge and spin transport on the surfaces of topological Nodal-line semimetals}
\label{sect2}
In our numerical calculations and analytical deductions, we mainly employ the non-equilibrium Green's function technique\cite{landauer1970,PhysRevLett.57.1761}. The two ends of the normal chain can be viewed as two terminals $L$ and $R$, and the part connected to the semimetal or the effective system is treated as the central scattering region. The zero-temperature differential charge conductance in the chain $G=d I / d V$ is connected to the transmission coefficient $T$ from terminal $L$ to $R$ via $G=\frac{e^2}{h} T$, where $T$ can be given by:
\begin{equation}
T=\text{Tr}\{\Gamma_L G^r \Gamma_R G^a\}.
\end{equation}
Here the trace is over the spin degree of freedom, and $G^r$ is the retarded Green's function which can be expressed as:
\begin{equation}
G^r(E)=(E-\mathbf{H}-\Sigma^r_L-\Sigma^r_R-\Sigma^r_S)^{-1},
\end{equation}
where $\mathbf{H}$ stands for the matrix Hamiltonian of the central scattering region and $\Sigma^r_i$ $(i=L,R,S)$ is the self-energy due to the coupling to terminal $L$, $R$, and the semimetal or the effective system. The diagonal matrix $\Gamma_{L/R}$ whose diagonal entries are proportional to the group velocities of the propagating eigenmodes, is the line-width function which can be calculated by $\Gamma_{L/R}=i[\Sigma_{L/R}^r-\Sigma_{L/R}^a]$. For simplicity, the 1D chain is chosen to be a normal metal, whose Hamiltonian is assumed to be
\begin{equation}
\mathcal{H}(k^N)=(-2 \cos k^N-\mu)I_{2\times2},
	\label{chain}
\end{equation}
where $\mu$ is the chemical potential. Thus according to Eq. (\ref{relationship}), the self-energy $\Sigma^r_{L/R}$ is found to be $\Sigma^r_{L}=\Sigma^r_{R}=-e^{ik^N}I_{2\times2}$, and $\Gamma_{L}=\Gamma_{R}=2\text{sin}k^N I_{2\times2}$. So the transmission coefficient $T$ can be simplified as $T=\sum_{\alpha,\beta=\uparrow,\downarrow}T_{\beta\alpha}$. Here $T_{\beta\alpha}=4\text{sin}^{2}k^N\times\mid G^r_{\beta\alpha}\mid^{2}$ represents spin-dependent transmission coefficient, where an incident electron with spin $\alpha$ from terminal $L$ is transmitted to terminal $R$ with spin $\beta$.

\subsection{1D chain coupled with a 1D effective system}
\begin{figure}[ht]
  \begin{center}
	\includegraphics[width=8.5cm,height=3.38cm]{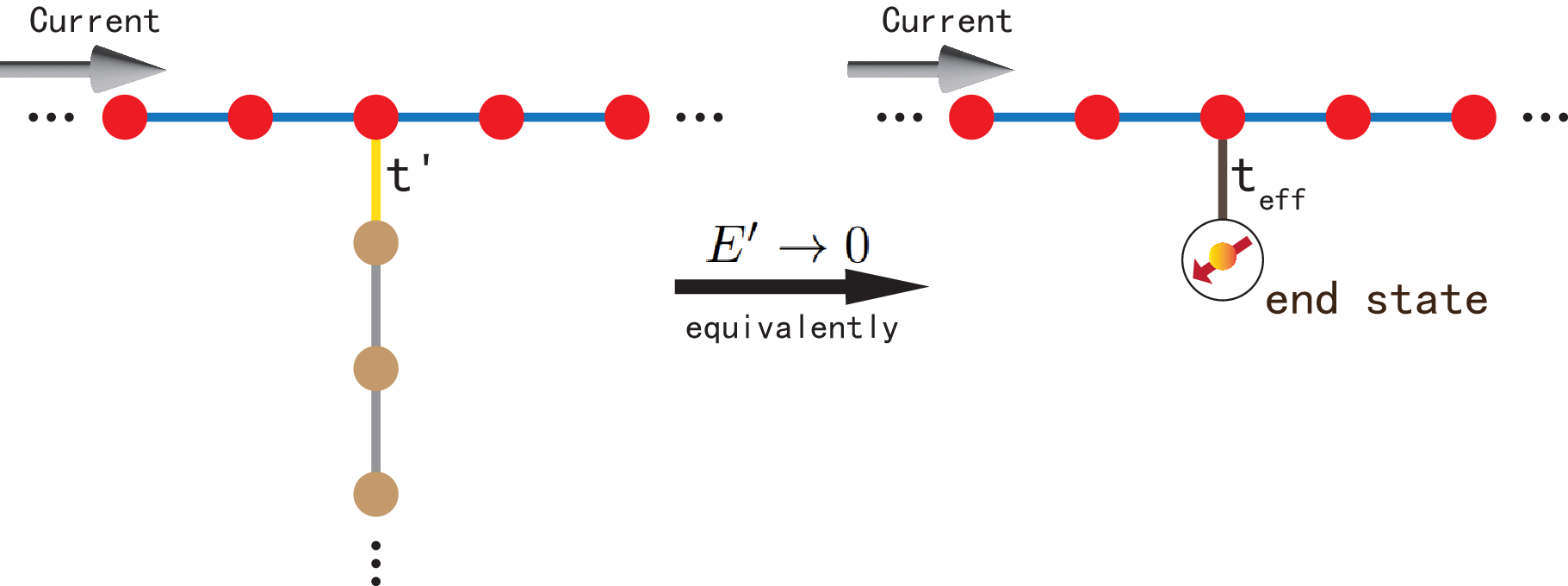}	
  \end{center}
  \vspace{-0.4cm}
  \caption{Current flowing in the metallic chain, which is coupled at one site to a 1D half-infinite effective system of a topological semimetal, described by Eq. (\ref{eq2}) (Left). The main physics of the quantum transport near the resonance ($E'\rightarrow0$) in the chain can be captured by considering that the chain is just coupled to a localized mode (Right).
  } \label{fig3}
\end{figure}

In this subsection, we focus on the following quantum-transport problem, where the metallic chain is coupled at one site with a half-infinite chain described by the 1D effective model of the topological semimetal, as schematically shown in Fig. \ref{fig3}. So we have $\Sigma^r_S (E)=t'^{2}g^r (E)$, where $t'$ is the coupling hopping integral between the metallic chain and the effective 1D system.

For the topological nontrivial case ($|\xi|<1$) with a positive $\lambda$, when $E'=E+\Delta\rightarrow 0$, according to Eq. (\ref{eq4}), the surface Green's function $g^r$ can be written as:
\begin{equation}
g^r(E')=a\{1-\sigma_y\}\frac{1}{E'^{+}}+b\sigma_x,
\label{eq8}
\end{equation}
where $a=\frac{2 \vert\lambda\vert (1-\xi^2)}{(1+|\lambda|)^2}$ and $b=-\frac{2\xi}{1+|\lambda|}$. Thus the retarded Green's function near the resonance can be derived as:
\begin{equation}
G^r(E'\to 0)=\frac{1+\sigma_y}{4 i \sin k^N},
\label{Gre}
\end{equation}
which is independent of $a$ and $b$. Then the transmission coefficient can be obtained as:
\begin{equation}
T(E' \to 0)=1.
\end{equation}

For clarity, in the following discussion of this paper we assume the spin polarization direction is always along $y$. As $E' \to 0$, we have $G^r_{\uparrow\uparrow}=(2i\text{sin}k^{N})^{-1}$ and $G^r_{\downarrow\downarrow}=G^r_{\uparrow\downarrow}=G^r_{\downarrow\uparrow}=0$. So we obtain $T_{\uparrow\uparrow}=1$ and $T_{\downarrow\downarrow}=T_{\uparrow\downarrow}=T_{\downarrow\uparrow}=0$. A remarkable fact is that the transmission current is fully spin polarized along $y$ when $E'=0$:
\begin{equation}
	P_y(E' \to 0)\equiv\frac{T_{\uparrow\uparrow}+T_{\uparrow\downarrow}-T_{\downarrow\downarrow}-T_{\downarrow\uparrow}}{T_{\uparrow\uparrow}+T_{\uparrow\downarrow}+T_{\downarrow\downarrow}+T_{\downarrow\uparrow}}=100\%.
\end{equation}

\begin{figure}[ht]
  \begin{center}
	\includegraphics[width=8.5cm,height=9.3cm]{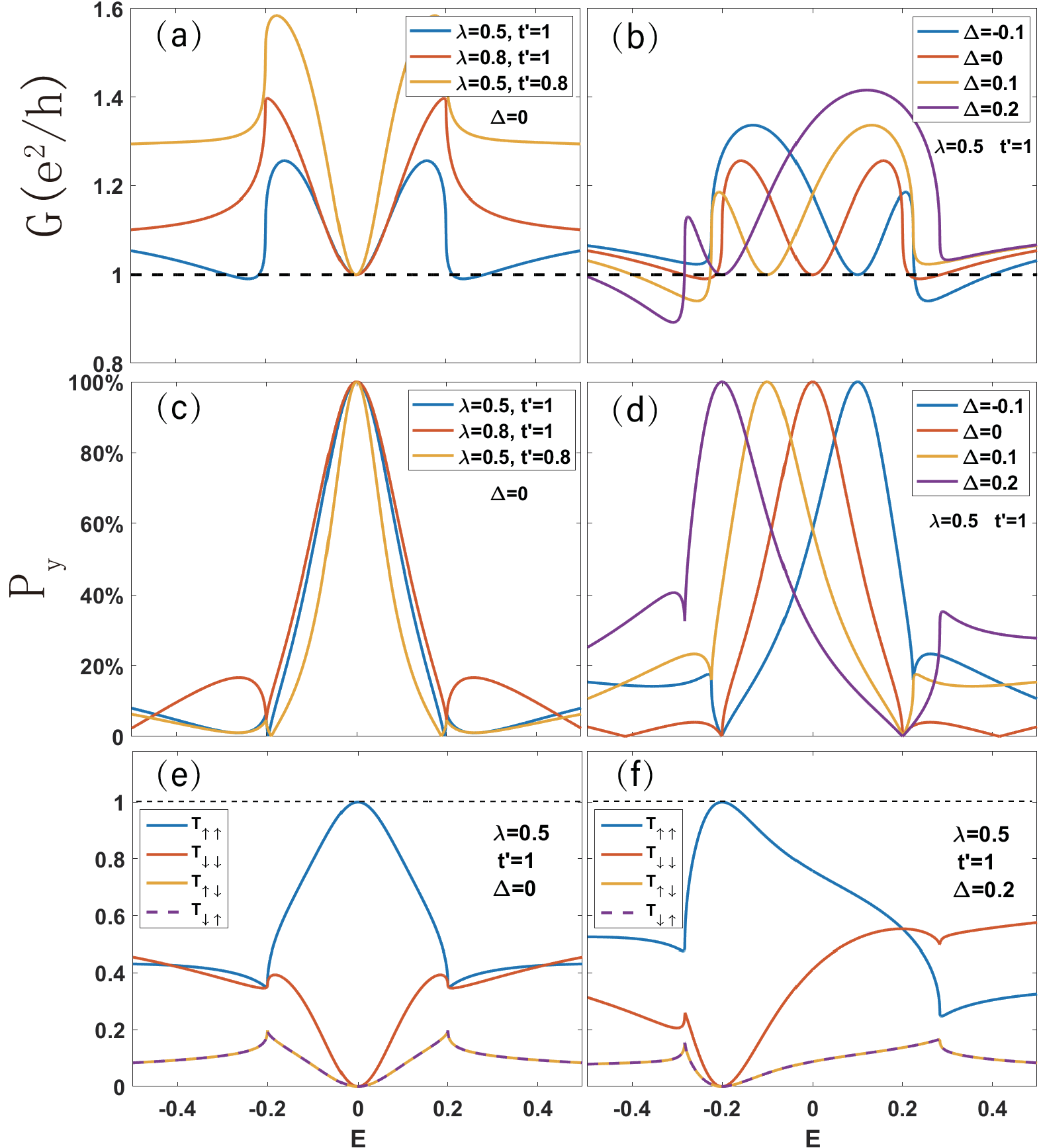}	
  \end{center}
  \vspace{-0.4cm}
	\caption{Transport properties of the metallic chain as functions of incident energy: 1D effective model case. (a)-(b) Charge conductance, (c)-(d)  current spin polarization and (e)-(f) spin-resolved transmission coefficients. Here $\mu=0$, $\xi=0.8$.
            }\label{fig4}
\end{figure}

The numerical results near the resonance energy $E'=0$ are shown in Fig. \ref{fig4}, where the charge conductance forms a valley at $e^{2}/h$ while the spin polarization $P_{y}$ forms a peak at $P_y=100\%$, no matter how the parameters of the model change. In addition, there is a symmetry for the curves: $T(E,\Delta)=T(-E,-\Delta)$ and $P_{y}(E,\Delta)=P_{y}(-E,-\Delta)$, i.e., only when $\Delta=0$, the curves are even functions of $E$.

This phenomenon can be physically explained as follows. The topologically nontrivial half-infinite chain has a localized spin-down end mode. Since $\Sigma^r_S (E)=t'^{2}g^r (E)$, according to Eq. (\ref{eq8}), near the resonance energy, the self-energy contribution from the effective system can be viewed as the metallic chain being coupled to an isolated site with on-site energy $E=-\Delta$. The effective coupling hopping matrix $t_{\text{eff}}$ between them is easily recognized to be:
\begin{equation}
t_{\text{eff}}=\sqrt{\frac{a}{2}} t'(1-\sigma_y)=\frac{\sqrt{\vert\lambda\vert (1-\xi^2)}}{1+|\lambda|}t'(1-\sigma_y).
\end{equation}
The isolated site is understood as the localized mode. This is schematically shown in Fig. \ref{fig3}. From this viewpoint, only the spin-down incident electron would be coupled to the isolated site, while the spin-up incident electron would be unaffected by the localized mode and would be totally transmitted to the right terminal, i.e., $T_{\uparrow\uparrow}=1$ and $T_{\downarrow\uparrow}=0$. On the other hand, the localized mode would cause the scattering rate to be infinite at the resonance so the incident spin-down electron would be totally reflected back to the left terminal. This means that $T_{\uparrow\downarrow}=T_{\downarrow\downarrow}=0$.

\subsection{1D chain on the edge of a 2D effective system}
\label{sect5}
\begin{figure}[ht]
  \begin{center}
	\includegraphics[width=8cm,height=4cm]{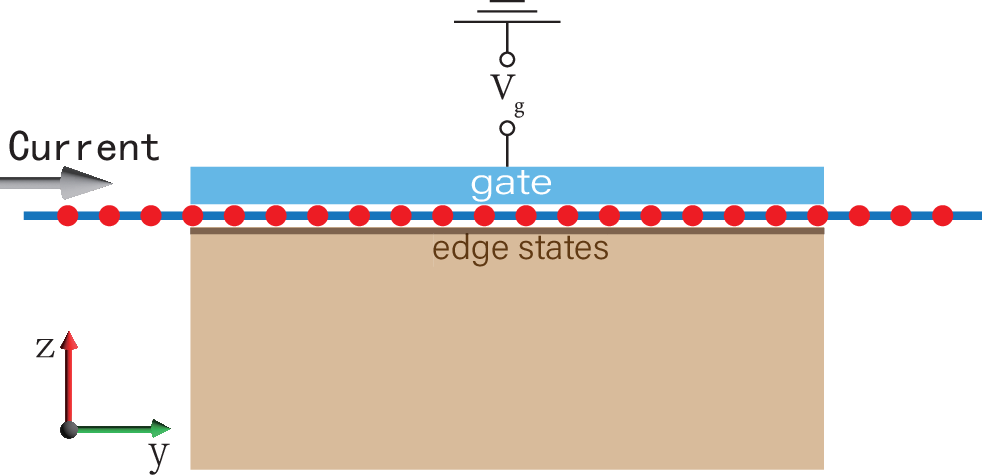}	
  \end{center}
  \vspace{-0.4cm}
	\caption{Current flowing in the metallic chain, which is coupled to the edge of a 2D half-infinite effective system of a topological nodal-line semimetal, described by Eq. (\ref{eq13}).
  } \label{fig5}
\end{figure}

In this subsection, as shown in Fig. \ref{fig5}, we study the following situation, where the 1D metallic chain is lying on the edge of a 2D effective system of the NLS described by Eq. (\ref{eq1}) with $\eta=0$, whose Hamiltonian is given by:
\begin{equation}
	\mathcal{H}_{\text{eff}} (k_y,k_z)=(m'-\cos k_y-\cos k_z)\sigma_x+ \lambda \sin k_z\sigma_z,
\label{eq13}
\end{equation}
where $m'=m-\cos k_x$, but here $m'$ should be understood as an independent parameter. This Hamiltonian describes a 2D Dirac semimetal with two Dirac nodes when $0<m'<2$. The two nodes are located at $\mathbf{k}=(k_y,k_z)=(\pm k_{0},0)$ in the momentum space, where $k_{0}=\cos^{-1}(m'-1)$. When the system is open along $z$ direction, there exists a flat edge band between the nodes projections: $\mid k_{y}\mid<k_{0}$. Here, to extract the main physics of the transport in the chain, a gate voltage $V_{g}$ is applied to slightly modify the energy of the surface modes, as shown in Fig. \ref{fig5}.

\begin{figure}[ht]
  \begin{center}
	\includegraphics[width=8.6cm,height=9cm]{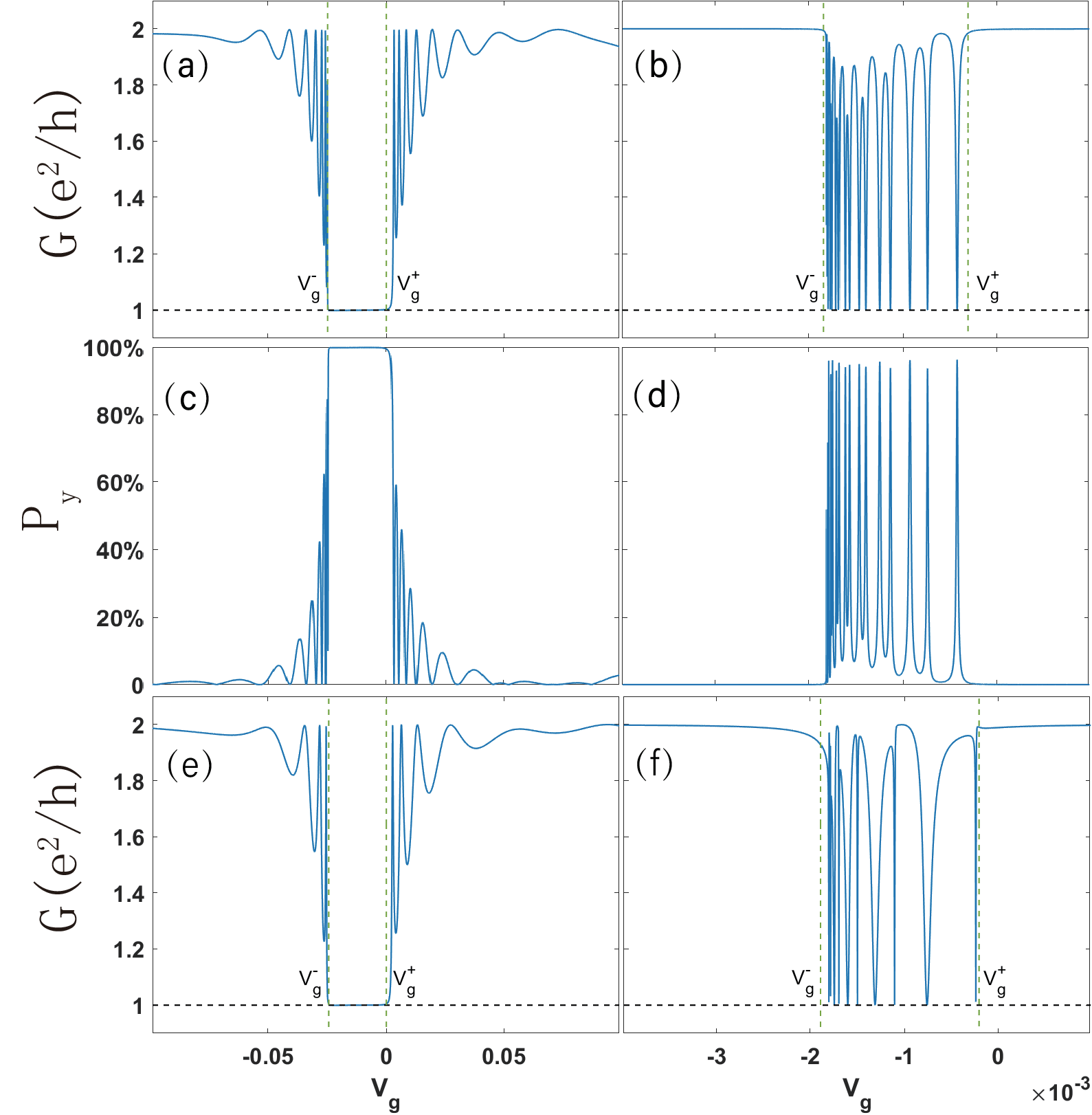}	
  \end{center}
  \vspace{-0.4cm}
	\caption{Transport properties of the metallic chain as functions of the gate voltage $V_g$: 2D effective model case. (a)-(b) Charge conductance and (c)-(d) spin polarization, where $t'=0.1$, $\lambda=0.5$, the chain length $L=50$ and $m'=1.5$ implying $k_{0}=\pi/3$. The incident wave vector $k^{N}$ is chosen to be $k^N=\pi/6<k_{0}$ for (a) and (c) while $k^N=5\pi/6>k_{0}$ for (b) and (d), which corresponds to that $\mu$ is chosen to be $\mu=\pm\sqrt{3}$ respectively. (e)-(f) The same as (a)-(b), except for $L=30$. The effective potential of the $n$th layer induced by the gate voltage is assumed to damp exponentially as $V_n=V_ge^{-n/5}$, where $n$ is the site index measured along $-y$ direction. Here the attenuation length is chosen to be $5$, and its variation does not qualitatively change the results.
  } \label{fig6}
\end{figure}

Now the self-energy $\Sigma^{r}_{S}$ is generically a function of parameter $m'$ and the width $L$ of the coupled chain: $\Sigma^{r}_{S}\equiv\Sigma^{r}_{S}(m',L)$. We have shown the numerical results of the quantum transport of the chain in Fig. \ref{fig6}. There are two distinct situations which depend on the incident wave vector $k^{N}$. As shown in Figs. \ref{fig6}(a), \ref{fig6}(b) and Figs. \ref{fig6}(e), \ref{fig6}(f), near $V_g=0$, if $k^{N}<k_{0}$, the transmission coefficient $T$ forms a plateau at $T=1$, while if $k^{N}>k_{0}$, $T$ forms lots of valleys at $T\approx1$. Correspondingly, at the same regime, the spin polarization $P_{y}$ of the transmitted current forms a plateau at $P_{y}=100\%$, as shown in Fig. \ref{fig6}(c), or $P_{y}$ forms multiple peaks at $P_{y}\approx100\%$, as shown in Fig. \ref{fig6}(d).

These results can be understood as follows. The central part of the chain which is coupled to the effective 2D system can be regarded as a 1D effective chain. The energy dispersion of this effective chain is corrected by the self-energy contributed from the coupling to the 2D effective system.  Assume the coupled chain is infinitely long, so $k$ is a good quantum number and then the self-energy becomes equivalently that from the coupling to the effective 1D system discussed above. Thus the effective Hamiltonian of the coupled chain can be given by:
\begin{equation}
h(k)=-2\cos k-\mu_{S}+t'^{2}g^r(E-V_{g}),
\end{equation}
where $\mu_{S}=\mu-V_{g}$ is the effective chemical potential of the coupled chain, and the $2\times2$ matrix $g^r(E)$ is the surface Green's function given by Eq. (\ref{eq4}) with $\xi=m'-\cos k$. So the resulting two energy bands can still be labeled by spin, namely, $\uparrow$ or $\downarrow$. For small coupling hopping $t'$, the spin-up band is nearly unchanged and its energy dispersion takes the form of the original metallic chain: $E_{\uparrow}\approx-2\cos k-\mu_{S}$. For the spin-down energy band, when $|k|>k_{0}$, which corresponds to the case of $\vert \xi \vert >1$, the correction from the self-energy is still negligible, because the surface Green's function $g^r(E)$ has no singularity. But when $|k|<k_{0}$, which corresponds to the case of $\vert \xi \vert <1$, the correction from the self-energy is large near the resonance, i.e., $E-V_{g}\rightarrow0$, and we obtain:
\begin{equation}
E_{\downarrow}\approx-2\cos k-\mu_{S}+\frac{2t'^{2}a(k)}{E-V_{g}},
\end{equation}
where
\begin{equation}
a(k)=\frac{2 \vert\lambda\vert (1-\xi^2)}{(1+|\lambda|)^2}=\frac{2 \vert\lambda\vert (1-(m'-\cos k)^2)}{(1+|\lambda|)^2}.
\end{equation}
Now we consider the scattering process. For the incident spin-down electron, we have $E_{\downarrow}=E=0$, so the above equation can be rewritten as:
\begin{equation}
\mu_{S}\approx-2\cos k-\frac{2t'^{2}a(k)}{V_{g}},
\label{eq17}
\end{equation}
where the right-hand part can be viewed as the transport associated energy band dispersion of the spin-down energy band.

\begin{figure}[ht]
	\begin{center}
	  \includegraphics[width=8.5cm,height=6.66cm]{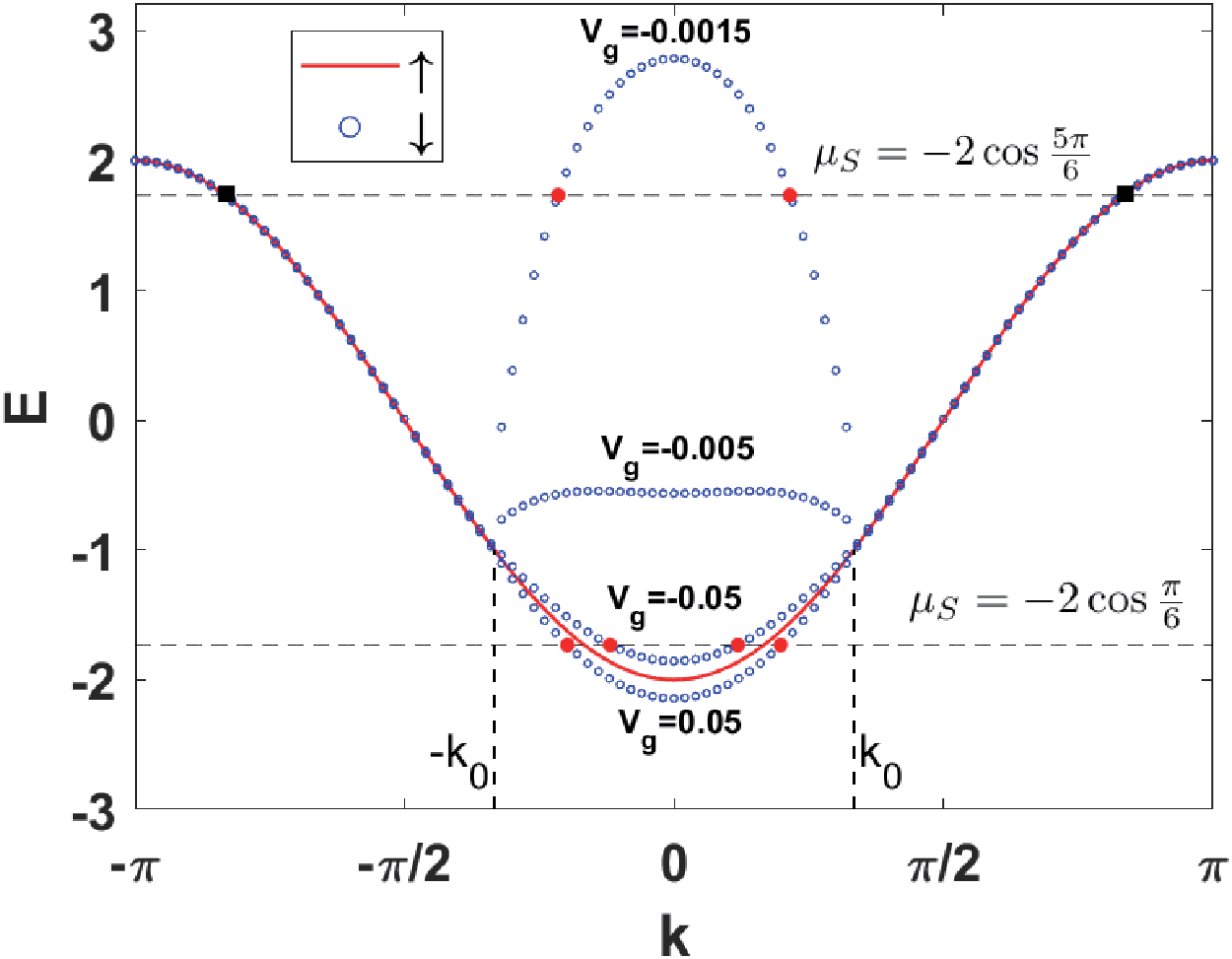}	
	\end{center}
	\vspace{-0.4cm}
	  \caption{Transport associated energy band dispersions of the 1D coupled chain for different gate voltage $V_{g}$, where the open circles (red line) represent the spin-down (-up) energy band. The solid circles and squares denote the propagating modes of the coupled chain at the corresponding chemical potential $\mu_{S}$ (horizontal dashed lines). Here $L=100$, $t'=0.1$, $\lambda=0.5$ and $m'=1.5$ implying $k_{0}=\pi/3$.
	} \label{fig7}
  \end{figure}

In Fig. \ref{fig7}, we show the exact numerical result of the modified energy dispersion for different gate voltage $V_{g}$. One can see that, the spin-up band is nearly unchanged compared with the original energy dispersion $-2\cos k$, which means the incident spin-up electron is nearly unaffected by the 2D substrate and would be nearly totally transmitted.

For the spin-down energy band, if the chemical potential $\mu_{S}$ is so small that $\mu_{S}<-2\cos k_{0}$, there always exists a window region of gate voltage $V_{g}$: $V^{-}_{g}<V_{g}<V^{+}_{g}\leq0$, within which there is no solution of $k$ to Eq. (\ref{eq17}). Here $V^{+}_{g}$ slightly depends on $L$, and as $L\rightarrow\infty$, $V^{+}_{g}\rightarrow0$, while $V^{-}_{g}\sim -t'^{2}$. This indicates that there is no propagating mode of spin-down electrons within the region, implying $T_{\uparrow\downarrow}=T_{\downarrow\downarrow}=0$. This results in the plateaus of charge conductance and current spin polarization as shown in Figs. \ref{fig6}(a), \ref{fig6}(c), and \ref{fig6}(e). When $V_{g}>V^{+}_{g}$ or $V_{g}<V^{-}_{g}$, there exist two solutions of $k$ to Eq. (\ref{eq17}): $\pm k^{*}$. Because $k^{*}<k_{0}$, these two states are expected to couple tightly with the edge modes of the 2D effective system. On the other hand, for a finite central part of the coupled chain with finite width $L$, the dispersion curves in Fig. \ref{fig7} are actually composed of discrete points (states) with small interval $\Delta k\approx2\pi/L$.  So as sweeping $V_{g}$, only when one of these isolated points lies on the line of $\mu_{S}$, can we obtain a solution. As a consequence, multiple valleys (peaks) appear on both outer sides of the window regions of the charge conductance (current spin polarization) in Figs. \ref{fig6}(a) and \ref{fig6}(e) (Fig. \ref{fig6}(c)). The total number of valleys(peaks) on either side can be estimated to be about $\frac{L}{\pi}\cos^{-1}(-\mu_{S}/2)$, or $\frac{L}{\pi}(k_{0}-\cos^{-1}(-\mu_{S}/2))$ respectively.

If $\mu_{S}>-2\cos k_{0}$, there are always two propagating states with their wave vectors being $\pm\cos^{-1}(-\mu_{S}/2)$ (black solid squares in Fig. \ref{fig7}). There still exists a window region of $V_{g}$: $V^{-}_{g}<V_{g}<V^{+}_{g}\leq0$, within which there are two additional solutions of $k$: $\pm k^{*}$ with $k^{*}<k_{0}$ (red solid dots in Fig. \ref{fig7}). The existence of these two new states indicates that they would be coupled tightly with the edge modes of the 2D effective system, indicating $T_{\uparrow\downarrow}\approx0$ and $T_{\downarrow\downarrow}\approx0$. Thus, there would be multiple valleys (peaks) within the window region of $V_{g}$ in the charge conductance (current spin polarization), as shown in Fig. \ref{fig6}(b) (Fig. \ref{fig6}(d)). The total number of valleys (peaks) is estimated to be about $\frac{k_{0}}{\pi}L$. However, when $V_{g}>V^{+}_{g}$ or $V_{g}<V^{-}_{g}$, there is no additional solution of $k$, which means the spin-down incident electron would be nearly not affected by the edge modes and so would be nearly totally transmitted, resulting in the plateaus at $2e^{2}/h$ for the charge conductance on both outer sides of the windows (See Fig. \ref{fig6}(b)).

\subsection{1D chain on the surface of a topological nodal-line semimetal}
\begin{figure*}[ht]
  \begin{center}
	\includegraphics[width=16cm,height=8.8cm]{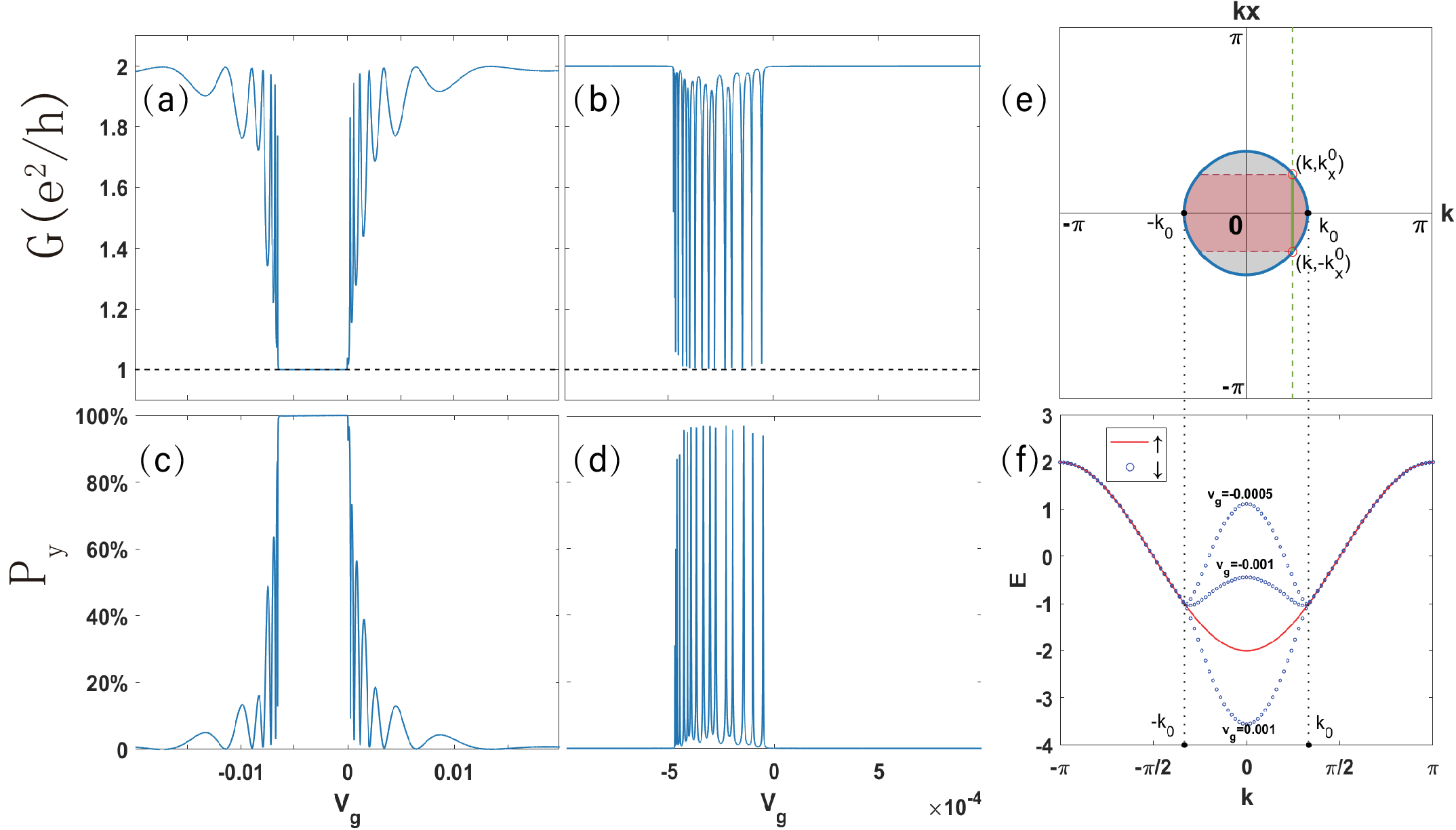}	
  \end{center}
  \vspace{-0.4cm}
	\caption{Transport properties of the metallic chain as functions of the gate voltage $V_g$: 3D topological nodal-line semimetal case, as shown schematically in Fig. \ref{fig1}. (a)-(b)  Charge conductance and (c)-(d) spin polarization, where $k^N=\pi/6$ for (a) and (c), while $k^N=5\pi/6$ for (b) and (d). (e)  Drumhead surface sates (grey region inside the blue circle), where for the effective coupled chain, if $k<k_{0}=\pi/3$ (green dashed line), the main self-energy contributions come from the $k_{x}$ characterized 2D effective systems with $\vert k_{x}\vert<k^{0}_{x}$ (red region restricted by the red dashed lines). (f) Transport associated energy band dispersions of the 1D coupled chain. Here parameters are set as $m=2.5$, $t'=0.1$, $\lambda=0.5$, and $L=50$ for (a)-(d) while $L=100$ for (f).
  } \label{fig8}
\end{figure*}
Lastly, we come to the most realistic situation: the transport of the chain on the surface of a topological NLS described by Eq. (\ref{eq1}) with $\eta=0$. On the top surface as shown in Fig. \ref{fig1}, there are drumhead surface states which are restricted within a loop: $\cos k_x+\cos k_y=m-1$, which is the projection of the nodal loop of the NLS on the surface. For simplicity, we adopt periodic boundary condition along $x$ direction, so $k_x$ is a good quantum number. In this case, because the deposited metallic chain is coupled only to the surface chain with $x=0$ of the NLS, the self-energy $\Sigma^r_S$ can be expressed as the sum over all those of the 2D effective system labeled by $k_{x}$ or $m'=m-\cos k_{x}$:
\begin{equation}
	\Sigma^r_S=\frac{1}{2\pi}\int^{\pi}_{-\pi}dk_{x}\Sigma^r_S(m-\cos k_{x},L),
\end{equation}
where the integrand is the self-energy contributed from the 2D effective system discussed in the last subsection. Our numerical results are shown in Figs. \ref{fig8}(a)-\ref{fig8}(d). The transport properties are also found to exhibit quite similar behaviors compared with those of the 2D case discussed above: there is a critical value of the incident wave vector $k^{N}$: $k_{0}$, which separates the two regions with distinct transport behavior. In this situation $k_{0}=\cos^{-1}(m-2)$.

These results can be understood by following the analogous arguments discussed in the last subsection. Consider an infinite 1D coupled chain, namely $L\rightarrow\infty$. Near the resonance energy, its effective Hamiltonian can be written as:
\begin{equation}
	h(k)=-2\cos k-\mu_{S}+\frac{t'^{2}}{2\pi}\int^{\pi}_{-\pi}dk_{x}g^r_S(E-V_{g}),
	\label{eq19}
\end{equation}
where $g^r_{S}(E)$ is the surface Green's function given by Eq. (\ref{eq4}), and $g^r_{S}(E)$ depends on $k$ and $k_{x}$ via $\xi=m-\cos k_x-\cos k$.
Similar to Eq. (\ref{eq17}), we have:
\begin{equation}
\mu_{S}\approx-2\cos k-\frac{t'^{2}}{\pi V_{g}}\int^{k^{0}_{x}}_{-k^{0}_{x}}dk_{x}a(k_{x},k),
\label{eq20}
\end{equation}
where the right-hand part represents the energy dispersion of the modified spin-down band, and,
\begin{equation}
a(k_{x},k)=\frac{2 \vert\lambda\vert}{(1+|\lambda|)^2}(1-(m-\cos k_x-\cos k)^2),
\end{equation}
with $k^{0}_{x}$ as a function of $k$ being under the constraint that the point $(k,k^{0}_{x})$ lies on the nodal loop of $\cos k+\cos k^{0}_x=m-1$. Here the spin-down band is modified mainly due to the coupling to the surface modes within $\vert k_{x}\vert<k^{0}_{x}$, as shown in Fig. \ref{fig8}(e), since all the drumhead surface states are spin-down localized modes. When $k>k_{0}=\cos^{-1}(m-2)$, there is no solution of $k^{0}_{x}$, which means the transport associated spin-down energy band becomes unaffected by the surface modes so that the critical value of $k$ is $k_{0}$. In Fig. \ref{fig8}(f), we give the numerical result of the modified energy dispersions. Because the spin-down band exhibits similar behavior compared to Fig. \ref{fig7}, the transport properties are expected to show analogous features.

\section{summary}
\label{sect3}
We have explored the transport properties of a metallic chain on the surface of a topological nodal-line semimetal. We first derived analytically the surface Green's function of the semimetal, which has a singularity as an indication of the existence of the surface modes. By making use of this result, we demonstrated that near the resonance, as function of the applied biased voltage, the charge conductance in the metallic chain shows a novel half-transmission behavior: a plateau at $e^2/h$ , while the spin polarization of the transmitted current reaches a plateau at $100\%$. This exotic phenomenon has been attributed to the effective coupling between the chain and the drumhead surface states, and so is well explained physically. This phenomenon is expected to serve as a transport signature of the Weyl nodal-line semimetals, and may also be helpful in detecting new semimetal materials in future.

\section{\textbf{Acknowledgment}}
This work is supported by NSFC under Grants No.11874202.

\begin{appendix}
\section{ANALYTICAL RESULTS OF THE SURFACE GREEN'S FUNCTION FOR THE 1D EFFECTIVE MODELS of TOPOLOGICAL SEMIMETALS}
\label{a1}
\begin{figure}[ht]
  \begin{center}
	\includegraphics[width=8.5cm,height=10.6cm]{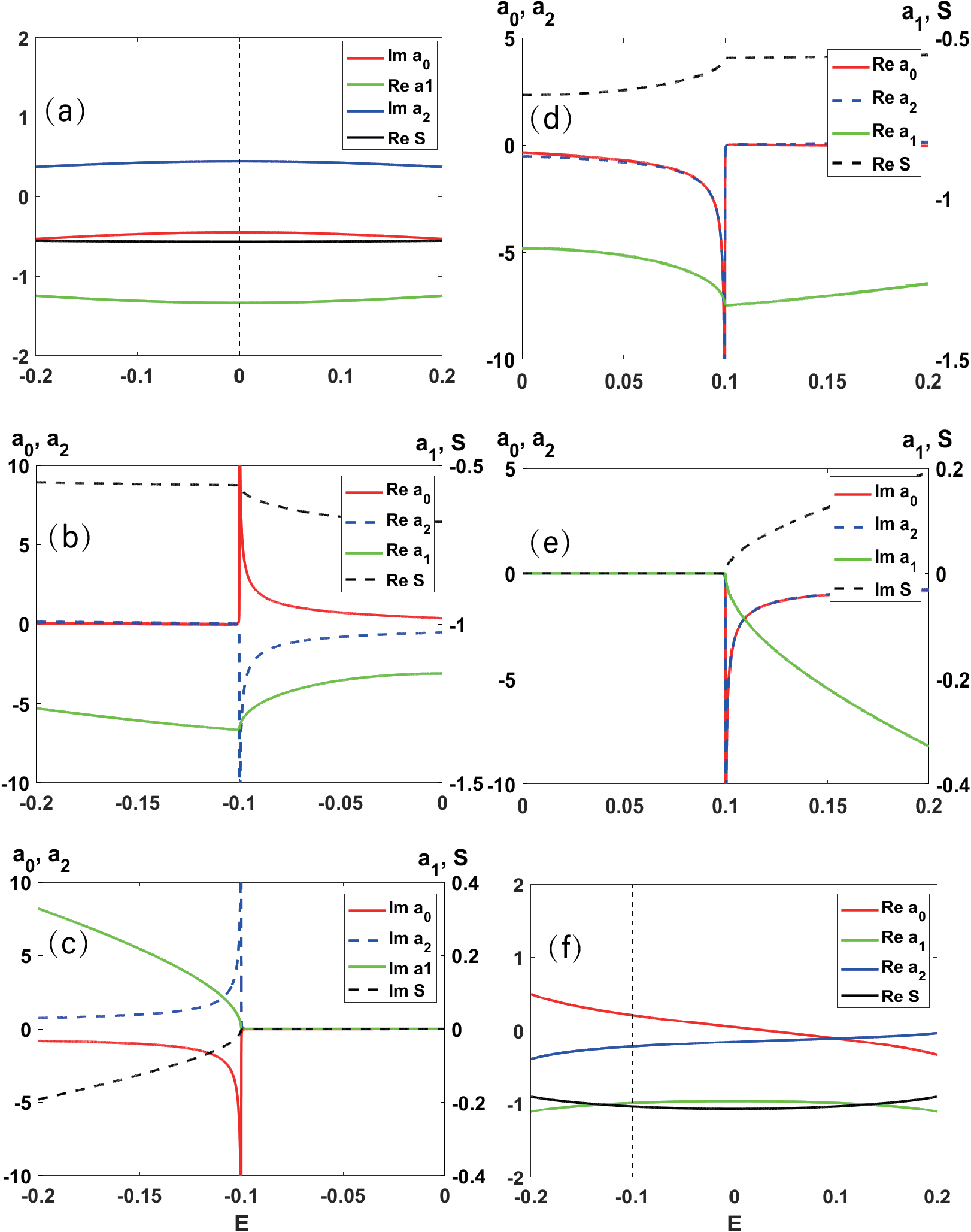}	
  \end{center}
  \vspace{-0.4cm}
	\caption {Numerical results for the expansion coefficients of the surface Green's function. (a) $\xi=1$, $\Delta=0$, $\lambda=0.5$. (b)-(e) $\xi=1$, $\Delta=0.1$, and $\lambda=0.5$ for (b)-(c), but $\lambda=-0.5$ for (d)-(e). (f) $\xi=1.2$, $\Delta=0.1$, $\lambda=0.5$.}
 \label{fig9}
\end{figure}

	In this appendix we give the detailed analytical derivation of the surface Green's function $g^r(E)$ for the 1D effective model Eq. (\ref{eq2}) of the topological semimetals. As a $2\times2$ matrix, $g^r$ can be expanded as $g^r=a_0+ \mathbf{a}\cdot \bm{\sigma}$. From Eq. (\ref{relationship}), the coefficients $a_0$ and $\mathbf{a}\equiv(a_{1},a_{2},a_{3})$ are found to satisfy:
\begin{align}
(S+\frac{1+\lambda^2}{4})a_0-\frac{\lambda}{2}a_2&=E^{+},
	\label{A1}\\
(S+\frac{1+\lambda^2}{4})a_2-\frac{\lambda}{2}a_0&=\Delta,
	\label{A2}\\
(S-\frac{1-\lambda^2}{4})a_1&=\xi,
    \label{A3}\\
a_3&=0
	\label{A4},
\end{align}
where $S=(a_0^2-\mathbf{a}^2)^{-1}$. It's hard to solve directly this set of equations. But by comparison with the corresponding numerical results as shown in the Fig. \ref{fig2} and Fig. \ref{fig9}, we can find the analytical expressions for $g^r(E)$ in the limit of $E\rightarrow-$sgn$(\lambda)\Delta$, namely, near the singularity of the surface Green's function. In the following we will derive the analytical expressions of the surface Green's function for the topologically nontrivial region $|\xi|<1$, the critical region $|\xi|=1$, and the topologically trivial region $|\xi|>1$ respectively.

First we consider the case of $|\xi|<1$. When $E\rightarrow-$sgn$(\lambda)\Delta$, the coefficients $a_{0}$ and $a_{2}$ are found to show singular behaviors. So we seek the following kind of solution near the singularity:
	\begin{align}
		&a_0=\frac{p_{-1}}{E'}+p_0+p_1E'+o(E'),
		\label{A5}\\
		&a_1=\frac{-2\xi}{1+|\lambda|},
        \label{A6}\\
		&a_2 =-\text{sgn}(\lambda)(\frac{p_{-1}}{E'}+p_0+q_1E')+o(E'),
		\label{A7}\\
		&S=-\frac{(1+|\lambda|)^2}{4}+\gamma_{1}E'+\gamma_{2} E'^2+o(E'^2),
  \label{A8}
     \end{align}
where $E'=E+$sgn$(\lambda)\Delta$, and $p_{0}$, $p_{\pm1}$, $q_{1}$, $\gamma_{1}$ and $\gamma_{2}$ are expansion coefficients. Because $a_{0}$ and $a_{2}$ are singular, the determinant of the coefficient matrix of Eq. (\ref{A1}) and Eq. (\ref{A2}) must be zero at $E'=0$. This is the reason why the leading term of $S$ in Eq. (\ref{A8}) takes the value of $-\frac{(1+|\lambda|)^2}{4}$. This then leads to Eq. (\ref{A6}) according to Eq. (\ref{A3}). Since $S^{-1}=a_0^2-a_1^2-a_2^2$, the finiteness of $S$ makes $a_{0}$ and $a_{2}$ share the same leading and subleading terms. According to Eqs. (\ref{A1})-(\ref{A2}), we have:
	\begin{equation}
		a_{0}+\text{sgn}(\lambda)a_{2}=-\frac{1}{\mid\lambda\mid}E'+o(E'),
		\label{A9}
	\end{equation}
which indicates $p_{1}-q_{1}=-1/\mid\lambda\mid$ and,
\begin{equation}
		a^{2}_{0}-a^{2}_{2}=\frac{-2p_{-1}}{\mid\lambda\mid}+o(1).
		\label{A10}
	\end{equation}
This together with Eq. (\ref{A3}) lead to:
	\begin{equation}
		p_{-1}=\frac{-\text{sgn}(\lambda)\Delta}{\gamma_{1}}=\frac{2|\lambda|(1-\xi^2)}{(1+|\lambda|)^2}.
  \label{A11}
	\end{equation}
 We remark that if $\Delta=0$, similar argument leads to the conclusion that $p_{-1}$ still takes the above expression but $\gamma_{1}=0$ in this case and $p_{-1}$ is related to $\gamma_{2}$ by $2\gamma_{2}p_{-1}=1$. So for the case of $|\xi|<1$, we get the surface Green's function up to the leading terms when $E\rightarrow -$$\text{sgn}$$(\lambda)\Delta$:
	\begin{equation}
\begin{split}
		&g^r(E)=\\
&\frac{2|\lambda|(1-\xi^2)}{(1+|\lambda|)^2}\{1-\text{sgn}(\lambda)\sigma_y\}\frac{1}{E+\text{sgn}(\lambda)\Delta}-\frac{2\xi}{1+|\lambda|}\sigma_x.
\end{split}
	\end{equation}
	
Secondly, we consider the critical case of $|\xi|=1$, when $E\rightarrow-$sgn$(\lambda)\Delta$. According to the numerical result as shown in Fig. \ref{fig9}(a), for $\Delta=0$, we seek the following kind of solution:
        \begin{align}
		&a_0 =p_0+p_1E'+o(E'),
  \label{A13}\\
		&a_1 =\frac{-2}{1+|\lambda|}(1+\frac{2\gamma_{1}}{1+|\lambda|}E')+o(E'),
  \label{A14}\\
		&a_2 =-\text{sgn}(\lambda)(p_0+q_{1}E'),
  \label{A15}\\
		&S =-\frac{(1+|\lambda|)^2}{4}+\gamma_{1} E'+o(E'),
  \label{A16}
	\end{align}
As before, the expression of $a_{1}$ in Eq. (\ref{A14}) is obtained by substituting $S$ in Eq. (\ref{A16}) into Eq. (\ref{A3}). Substituting the above equations into  $S^{-1}=a_0^2-a_1^2-a_2^2-a_3^2$, we get:
        \begin{equation}
		p_0(p_1-q_1)=\frac{8|\lambda|\gamma_{1}}{(1+|\lambda|)^4}.
  \label{A17}
	\end{equation}
From Eqs. (\ref{A1})-(\ref{A2}), we have:
        \begin{equation}
		\begin{split}
			a_0 &=-\text{sgn}(\lambda)\{1-\frac{2\gamma_{1}}{|\lambda|}E'+o(E')\}a_2+o(E')\\
			    &=p_0+(q_1-\frac{2p_0\gamma_{1}}{|\lambda|})E'+o(E').
       \label{A18}
		\end{split}
	 \end{equation}
Comparing the coefficient of the $E'$ term with Eq. (\ref{A13}), we get:
        \begin{equation}
		p_1-q_1=-\frac{2p_0\gamma_{1}}{|\lambda|}
		\label{A19}.
	\end{equation}
Similarly, we further have:
        \begin{equation}
		a_2=-\text{sgn}(\lambda)\frac{1}{2\gamma_{1}}+o(1),
			\label{A20}
	\end{equation}
which means that $p_0=\frac{1}{2\gamma_{1}}$. So $p_1-q_1=-\frac{1}{|\lambda|}$. Therefore we get:
        \begin{equation}
        p_0=\frac{-2|\lambda|i}{(1+|\lambda|)^2},
        \label{21}
	\end{equation}
and we obtain for $\Delta=0$, $\xi=1$ the surface Green's function up to the leading terms when $E\rightarrow0$:
        \begin{equation}
		g^r(E)=\frac{-2|\lambda|i}{(1+|\lambda|)^2}\{1-\text{sgn}(\lambda)\sigma_y\}-\frac{2}{1+|\lambda|}\sigma_x.
	\end{equation}
For $\Delta\neq0$, we seek the following kind of solution according to Figs. \ref{fig9}(b)-\ref{fig9}(e):
        \begin{align}
		&a_0 =\frac{p_{-\frac{1}{2}}}{\sqrt{E'}}+p_{0}+p_{\frac{1}{2}}\sqrt{E'}+o(\sqrt{E'}),
  \label{A23}\\
		&a_1 =\frac{-2}{1+|\lambda|}(1+\frac{2\gamma_{\frac{1}{2}}}{1+|\lambda|}\sqrt{E'})+o(\sqrt{E'}),
  \label{A24}\\
		&a_2 =-\text{sgn}(\lambda)a_{0}+o(\sqrt{E'}),
  \label{A25}\\
		&S =-\frac{(1+|\lambda|)^2}{4}+\gamma_{\frac{1}{2}}\sqrt{E'}+o(\sqrt{E'}).
  \label{A26}
	\end{align}
Eq. (\ref{A9}) is still true, so we have:
\begin{align}
        a^{2}_{0}-a^{2}_{2}&=\frac{2\Delta}{\lambda\gamma_{\frac{1}{2}}}\sqrt{E'}+o(\sqrt{E'}),
        \label{A27}\\
        p_{-\frac{1}{2}}&=-\frac{\text{sgn}(\lambda)\Delta}{\gamma_{\frac{1}{2}}}.
        \label{A28}
\end{align}
Substituting Eq. (\ref{A28}) and Eq. (\ref{A24}) into the expression of $S^{-1}$, we get:
\begin{equation}
        \gamma^{2}_{\frac{1}{2}}=\frac{\text{sgn}(\lambda)\Delta(1+\mid\lambda\mid)^{4}}{8\lambda^{2}}.
        \label{A29}
\end{equation}

So for $\Delta\neq0$ and $\xi=1$, we have the surface Green's function up to the leading terms when $E\rightarrow-\text{sgn}(\lambda)\Delta$:
        \begin{equation}
		g^r(E)=\frac{p_{-\frac{1}{2}}}{\sqrt{E'}}\{1-\text{sgn}(\lambda)\sigma_y\}-\frac{2}{1+|\lambda|}\sigma_x.
	\end{equation}
where
\begin{align}
        p_{-\frac{1}{2}}&=\frac{\sqrt{8\vert\Delta\vert}\mid\lambda\mid}{(1+\mid\lambda\mid)^{2}}, \quad\quad  \text{if} \quad \lambda\Delta>0,\\
        p_{-\frac{1}{2}}&=-i\frac{\sqrt{8\vert\Delta\vert}\mid\lambda\mid}{(1+\mid\lambda\mid)^{2}}, \quad \text{if} \quad \lambda\Delta<0.
\end{align}
Finally, we consider the topologically trivial case of $|\xi|>1$, when $E\rightarrow-$sgn$(\lambda)\Delta$. According to the numerical result as shown in Fig. \ref{fig9}(f), we seek the following kind of solution:
\begin{align}
		a_0 &=p_{0}+p_{1}E'+o(E'),
  \label{A33}\\
		a_1 &=\frac{\xi}{S-\frac{1-\lambda^2}{4}},
  \label{A34}\\
		a_2 &=-\text{sgn}(\lambda)p_{0}+q_{1}E'+o(E').
  \label{A35}\\
\end{align}
Since $S^{-1}=a_0^2-a_1^2-a_2^2$, we find
\begin{equation}
		S=-a^{-2}_{1}=-\frac{1}{4}(\xi+\sqrt{\xi^2+\lambda^2-1})^2.
  \label{A37}
\end{equation}
From Eqs. (\ref{A1})-(\ref{A2}), the leading term's coefficient $p_{0}$ is found to be:
\begin{equation}
p_{0}=-\frac{\text{sgn}(\lambda)\Delta}{S+(1+|\lambda|)^{2}/4},
\end{equation}
$p_1$ and $q_1$ can then be further determined as:
\begin{equation}
p_{1}=\frac{S+(1+\lambda^2)/4}{(1-\xi^2)S}, \quad q_{1}=\frac{\lambda}{2(1-\xi^2)S}.
\end{equation}
So for $\xi>1$ we get the surface Green's function up to the leading terms in this case:
  \begin{equation}
\begin{split}
	g^r(E) &=\frac{\Delta}{S+(1+|\lambda|)^{2}/4}\{-\text{sgn}(\lambda)+\sigma_{y}\}-\frac{1}{\sqrt{-S}}\sigma_{x}\\
&+\frac{1}{(1-\xi^2)S}\{(S+\frac{1+\lambda^2}{4})+\frac{\lambda}{2}\sigma_y\}E'.
        \end{split}
\end{equation}
Note that if $\Delta\neq0$, the first term is the leading term, otherwise, the third term is the corresponding leading term instead.
\end{appendix}

\bibliographystyle{unsrt}
\bibliography{nodal}
\end{document}